\documentstyle[psfig]{l-aa}         % referee style: [psfig,referee]{l-aa}

\def\ros{{\sl ROSAT }}
\def\ein{{\sl Einstein }}
\def\asca{{\sl ASCA }}
\def\ginga{{\sl Ginga }}
\def\exo{{\sl EXOSAT }} 
\def\heao1{{\sl HEAO 1 }}

\def\G{$\Gamma_{\rm x}$ } 

\def\degs{\ifmmode ^{\circ}\else$^{\circ}$\fi}
\def\amin{\ifmmode ^{\prime}\else$^{\prime}$\fi}
\def\asec{\ifmmode ^{\prime\prime}\else$^{\prime\prime}$\fi}
\def\approxlt{\mathrel{\hbox{\rlap{\lower.55ex \hbox {$\sim$}}
        \kern-.3em \raise.4ex \hbox{$<$}}}}
\def\approxgt{\mathrel{\hbox{\rlap{\lower.55ex \hbox {$\sim$}}
        \kern-.3em \raise.4ex \hbox{$>$}}}}

\begin{document}
 
   \thesaurus{03         % A&A Section 3: Extragalactic Astronomy
              (11.01.2;  % Galaxies: active
               11.09.1;  % Galaxies: individual  
               11.17.2;  % Galaxies: emission lines 
               11.19.1;  % Galaxies: Seyfert
               13.25.2)  % X-rays: galaxies
}

   \title{Evidence for a {\em dusty} warm absorber in NGC 3227 ?} 
   \author{Stefanie Komossa, Henner Fink$^{\dagger}$} 

   \offprints{St. Komossa, skomossa@mpe-garching.mpg.de  \\
       $\dagger$ deceased in Dec. 1996}
 
  \institute{Max-Planck-Institut f\"ur extraterrestrische Physik,
             85740 Garching, Germany\\
       }

   \date{Received 11 December 1996; accepted 19 June 1997}

   \maketitle\markboth{St. Komossa, H. Fink~~~ Dusty warm absorber in NGC 3227}{}  

   \begin{abstract}
%______________________________________ Do not leave a blank line here!
We have analyzed survey and pointed \ros PSPC observations of the
Seyfert galaxy NGC 3227.
Large amplitude X-ray variability is detected, with a factor $\sim$ 15 change
in count rate within about 3 years. Smaller changes are seen on the timescale
of days,
the largest being a factor of 3.5.  
No strong {\em spectral} variability is found throughout the pointed observation.
The X-ray spectrum is modeled in terms of warm absorption
and both, a dust-free warm absorber and one with internal dust,  
give an excellent spectral fit.  
Additionally using multi-wavelength observations of NGC 3227, 
the {\em dusty} warm absorber is favored and  
suggested as an explanation for the lack of a large cold absorbing
column in the soft X-ray region despite excess reddening along the line of sight. 
Self-consistent re-calculation of the models including Galactic-ISM-like dust 
results in an ionization parameter $\log U \simeq -0.25$ and 
a column density $\log N_{\rm w} \simeq 21.8$ of the warm absorber.  
X-ray variability and dust survival arguments are used to constrain
the density and further properties of the ionized material. 
The influence of dust and other parameters on the thermal stability
of the warm absorber is investigated.      
Predictions are given for the absorber-intrinsic optical--UV line 
emission and absorption.
 

   \end{abstract}
 
\section{Introduction}

NGC 3227 is a Seyfert 1.5 galaxy 
at a redshift of $z$=0.003. It shows
signs of interaction with its dwarf elliptical companion NGC 3226. 
Although studied in many spectral regions,  
most attention has focussed on the optical wavelength range:  
NGC 3227 has been the target of
several BLR mapping campaigns (Rosenblatt et al. 1992, Salamanca et al. 1994, Winge et al. 1995)
and high-resolution line-profile studies (e.g. Whittle 1985, Rosenblatt et al. 1994).
High spatial resolution observations of the kinematics and ionization
structure of the circum-nuclear region were presented by 
Mediavilla \& Arribas (1993) and Arribas 
\& Mediavilla (1994).  
Rubin \& Ford (1968) and Mundell et al. (1995a) reported evidence for very high emission-line reddening. 

X-ray variability of NGC 3227 was detected with \heao1   
by Tennant \& Mushotzky (1983), who found 
the flux to increase by a factor of 1.4 within 5 hours.
Variability was also found in the 
\exo data (Turner \& Pounds 1989)   
and in an \asca observation (Ptak et al. 1994).   

Reichert et al. (1985), using \ein data, favored a partial covering
model to describe the (0.8--4.5) keV X-ray spectrum of NGC 3227. 
Differential variability between the \exo LE (0.05--2 keV) and ME (2--10 keV)  
instruments was 
interpreted in terms of either partial covering with a strongly varying
cold column  
or a variable soft excess (Turner \& Pounds 1989).  
An unusually flat powerlaw of photon index \G $\approx$ --1.5 was obtained.  
An iron K-emission line near 6.5 keV was detected in the \ginga 
spectrum of NGC 3227 (Pounds et al. 1989).      
This observation was modeled by George et al. (1990) in terms
of reflection from a cold accretion disk. The underlying X-ray 
continuum emission was then best described by a powerlaw with
index \G = --1.86$\pm$0.04. 
Given the lack of spectral resolution below $\sim$ 1 keV 
in these studies, it was not possible to distinguish between
alternative descriptions of the soft spectral shape and
to evaluate the cause of variability.

Evidence for the presence of a warm absorber in NGC 3227 
was indicated by Netzer et al. (1994) by the use of X-ray color diagrams, and  
reported by Ptak et al. (1994) in an
\asca observation.      
As the cause for variability detected in the \asca data,
Ptak et al. found two possibilities. One is a change in powerlaw index by $\Delta$\G $\approx$ 0.2
and constant properties of the ionized material. The other is
complex behavior of the warm absorber with variability of both, ionization parameter 
and warm column density.  
  
For the present study, we have analyzed the survey and archival pointed \ros 
observations of NGC 3227  
performed with the PSPC
(Tr\"umper 1983; Pfeffermann et al. 1987).   
Since there is non-X-ray evidence for a large cold column of absorbing material along
the line of sight (e.g. Cohen 1983), and as the absorption structure of
the comparatively lowly ionized warm absorber extends well below 0.5 keV,  
\ros is particularly well suited for an investigation of these features. 
Furthermore, in 
a sequence of studies to assess 
the possibility of a warm-absorber contribution to 
one of the known high-ionization emission-line regions in individual AGN (Komossa \& Fink, e.g. 1997b),  
NGC 3227 marks the low-ionization end of the observed warm absorbers.
In this context, it is also interesting to note that Salamanca et al. (1994)
reported the existence of broad wings in H$\alpha$ in NGC 3227 that do not
follow the continuum variability, indicating a higher than usual degree
of ionization, as would be expected for a warm absorber.   

The paper is organized as follows: 
In Sect. 2 we present the observations. 
The data are analyzed with respect to their spectral and temporal properties in 
Sects. 3 and 4, respectively. 
In Sect. 5, evidence for the presence of dust mixed with the warm absorber 
is presented and compared to alternative scenarios. Further properties of the ionized material are derived and 
discussed, and the absorber-intrinsic line emission and absorption in the optical--UV 
spectral region is studied.   
A summary and the conclusions are provided in Sect. 6. 

A distance of 18 Mpc is adopted for NGC 3227 assuming a Hubble constant of $H_{\rm o}$ = 50 km/s/Mpc
and the galaxy to follow the Hubble flow.  
 
If not stated otherwise, cgs units are used throughout.

\section{Data reduction}

\subsection{Pointed data} 
The observation was performed with the \ros PSPC from May 8 - 19, 1993, 
centered on NGC 3227.   
The total exposure time is 19.6 ksec.   
The source photons were extracted 
within a circle chosen to be large enough to ensure that all of the source counts 
were included, but small enough to contain negligible contribution from
the much weaker neighboring source NGC 3226.  
The background was determined after removing all detected sources within the inner 
19' of the field of view.
The data were corrected for 
vignetting
and dead-time using the EXSAS software package (Zimmermann et al. 1994).
The mean source count rate is about 0.5 cts/s.
For the spectral analysis source photons in
the amplitude channels 11-240 were binned 
according to a constant signal/noise ratio of 21$\sigma$. 
For the temporal analysis the minimal bin size in time was 400 s to account for the
wobble mode of the observation. 

\subsection{Survey data}
 NGC 3227 was observed during the \ros all-sky survey (RASS)
from Nov. 12 -- 15, 1990, with an effective exposure time of 610 s.
It is clearly weaker in this observation as compared to the pointing.  
However, due to the larger effective pointspread function
during the survey observations, it was not possible to exclude the
contribution of the weak neighboring source NGC 3226 (although the position of the centroid of the X-ray emission 
agrees with the coordinates of NGC 3227). Therefore, `source'photons 
within a circle including both objects were extracted. The background was determined 
from two source free sky fields along the scanning direction of the telescope. 
 The data were corrected for vignetting.   
A total number of 52 source photons was detected. Assuming NGC 3226 to have remained 
constant in count rate from the RASS to the pointed observation (with 0.018 cts/s),
results in a count rate of 0.067 cts/s for NGC 3227, revealing strong variability.   
Unfortunately, the low number of photons accumulated during the RASS does not 
allow a more detailed discussion
of the spectral and temporal properties. 

\section{Spectral analysis} 

\subsection{Standard spectral models}
A single powerlaw (with cold absorption column as a free parameter)
is a bad description of the soft X-ray spectrum 
($\chi{^{2}}_{\rm red}$=1.7; Table 1).
The resulting powerlaw is very {\em flat}, with an index $\Gamma_{\rm{x}}$=--1.2, 
and strong systematic residuals remain (Fig. \ref{SEDx}). 
None of several single-component models that were compared with the data 
provides an acceptable fit. This also holds for 2-component (powerlaw plus soft excess) models 
in which 
the  powerlaw index \G is fixed to --1.9 with all other parameters free.  

A powerlaw of free index with a soft excess, 
or a warm-absorbed powerlaw of canonical index (next section) give successful fits. 
The soft excess was parameterized as black body, for which we find $kT_{\rm bb} \simeq$ 0.07 keV,
or the standard accretion disk model after Shakura \& Sunyaev (1973). The latter yields
an unusually low black hole mass of $M_{\rm BH} \simeq$ 3 $\times 10^{4}$ M$_{\odot}$ for fixed 
accretion rate ${\dot M}/{\dot M_{\rm edd}}$ = 1, and a cold column density of 
$N_{\rm H} \simeq 0.70 \times 10^{21}$ cm$^{-2}$ (Table 1).   
Since also  
the underlying powerlaw is again unrealistically flat, with $\Gamma_{\rm{x}} \simeq$ --1.1,  
in contradiction to the higher-energy observations (e.g. George et al. 1990, 
$\Gamma_{\rm{x}} \simeq$ --1.9), 
this model is not discussed further.

 Finally, we applied a partial covering model to the data, in which part of an intrinsic 
powerlaw spectrum is directly seen, and part is absorbed by a very high cold column
($N_{\rm H}$ left free or fixed to 8.3 $\times 10^{21}$ cm$^{-2}$; cf. Table 2). 
This approach is motivated by the evidence of large amounts of cold matter along
the line of sight deduced from emission-line reddening, as is further discussed below.   
No successful description of the data was achieved.  

\subsection{Warm absorber models}  

\subsubsection{Model properties and assumptions} 

%%\begin{figure}
%%\picplace{6cm}
  \begin{figure}[thbp]
      \vbox{\psfig{figure=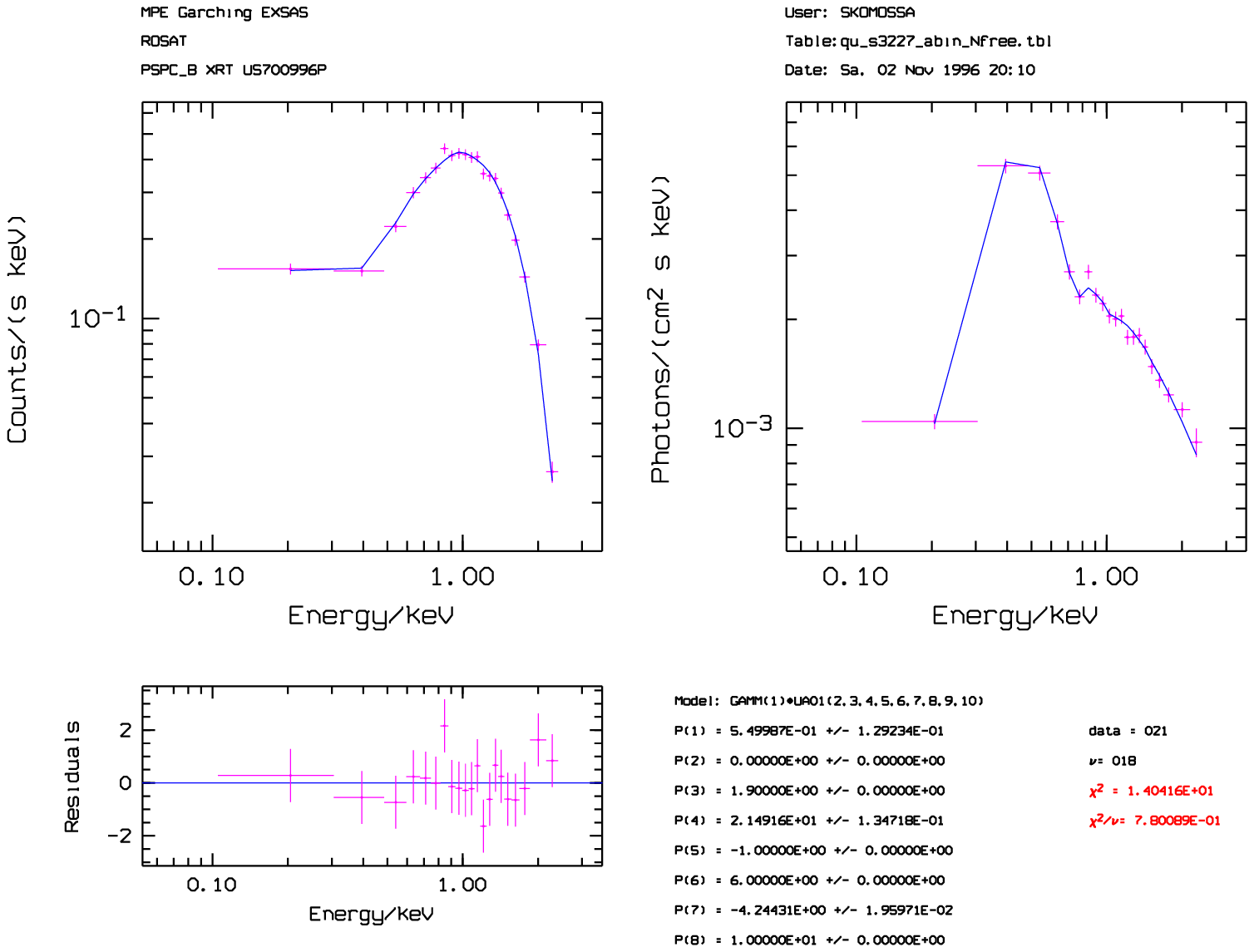,width=8.8cm,%
          bbllx=2.5cm,bblly=1.1cm,bburx=10.1cm,bbury=11.7cm,clip=}}\par
            \vspace{-0.7cm}
      \vbox{\psfig{figure=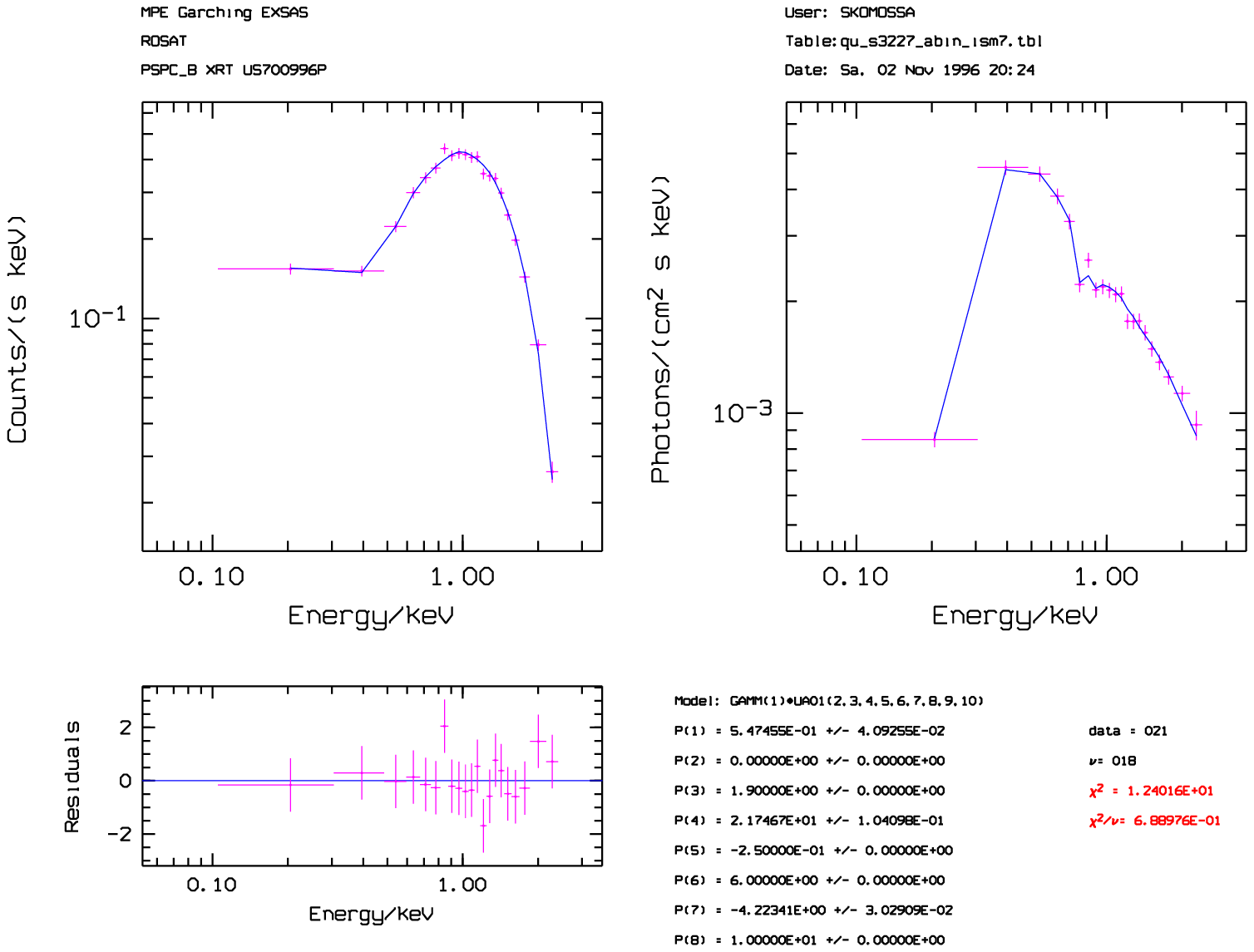,width=8.8cm,%
          bbllx=2.5cm,bblly=1.1cm,bburx=10.1cm,bbury=4.5cm,clip=}}\par
            \vspace{-0.7cm}
      \vbox{\psfig{figure=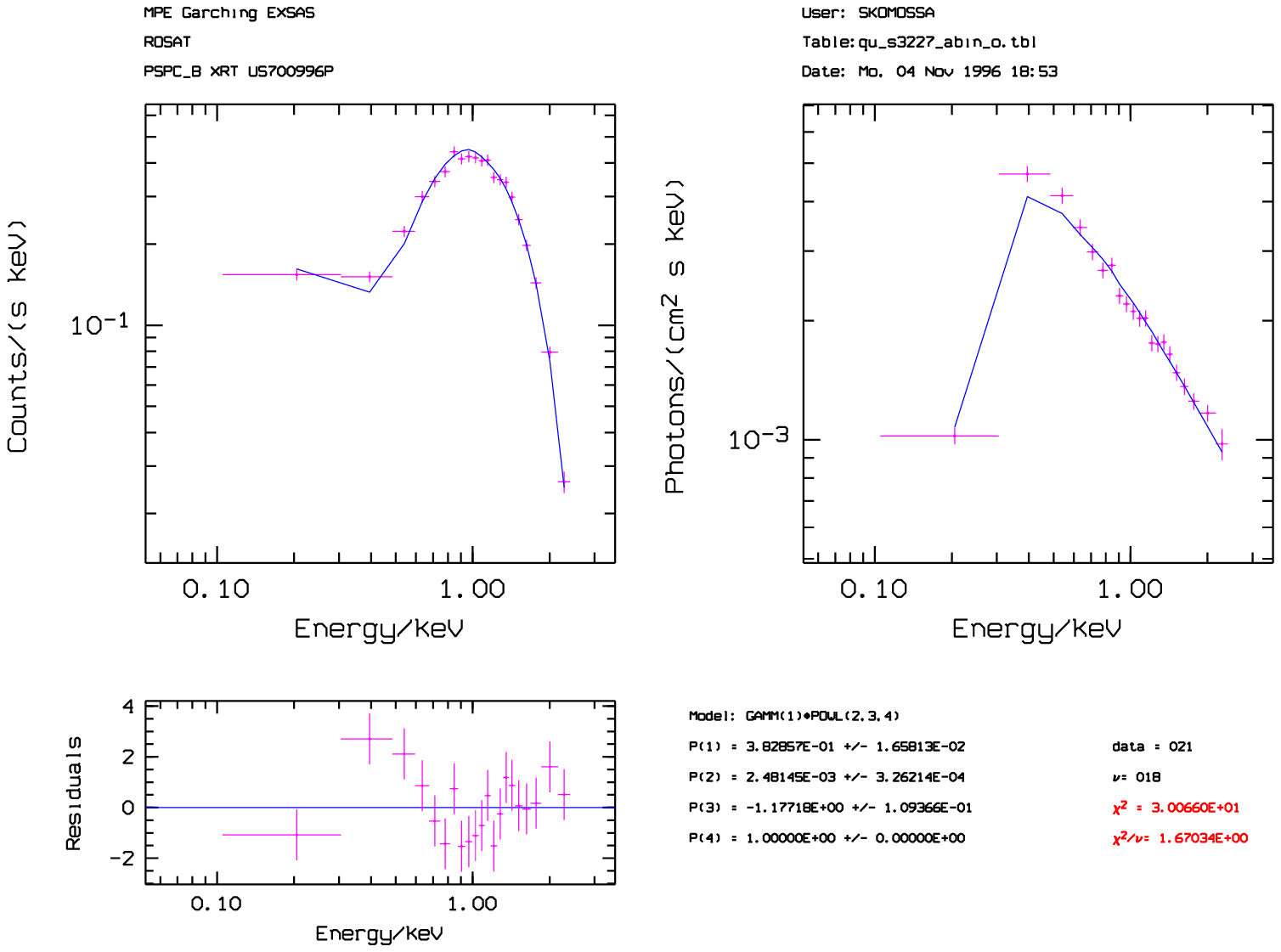,width=8.8cm,%
          bbllx=2.5cm,bblly=1.1cm,bburx=10.1cm,bbury=4.5cm,clip=}}\par
%            \vspace{-0.4cm}
\caption[SEDx]{ The upper panel shows the observed X-ray spectrum of NGC 3227 (crosses)
and the best-fit dust-free warm absorber model (solid line). The second panel displays
the fit residuals for this model, and the third panel the residuals from the 
best-fit dusty warm absorber. For comparison, the residuals resulting from
a single powerlaw description of the data are shown in the lowest panel
(note the different scale in the ordinate). 
  }
\label{SEDx}
\end{figure}
To model the X-ray spectrum in terms of warm absorption, we have 
calculated a sequence of photoionization models
for the warm material, using the code {\em Cloudy} (Ferland 1993).
We assume the material (a) to be photoionized by continuum emission of the central
pointlike nucleus, (b) to be one-component and of constant density, 
and (c) to have solar abundances (Grevesse \& Anders 1989). 
In those models that include dust mixed with the warm gas,
dust properties 
like those of the Galactic diffuse interstellar medium
are adopted (Mathis, Rumpl \& Nordsieck 1977; MRN) if not mentioned otherwise, 
and the gas-phase abundances are depleted correspondingly (a mean of Cowie \& Songaila 1986), 
as in Ferland (1993). The MRN dust consists of two grain species, graphite and
astronomical silicate, and a powerlaw grain size distribution. 
     
The ionization state of the warm absorber can be characterized by the
hydrogen column density $N_{\rm w}$ of the ionized material and the
ionization parameter $U$, defined as  
\begin {equation}
U=Q/(4\pi{r}^{2}n_{\rm H}c)
\end {equation} 
where $Q$ is the
number rate of incident photons above the Lyman limit, $r$ is the distance between
nucleus and warm absorber, $n_{\rm H}$ is the hydrogen density 
(fixed to 10$^{9.5}$ cm$^{-3}$ unless noted otherwise) 
and $c$ the speed of light.  Both quantities, $N_{\rm w}$ and $U$, are determined from the X-ray
spectral fits. 
 
The spectral energy distribution (SED) chosen for the mo\-deling corresponds to  
a `mean Seyfert' continuum consisting of (i) piecewise powerlaws from the radio 
to the gamma-ray spectral region, or (ii) an additional EUV bump parameterized as a black 
body of $T$ = 120\,000 K contributing the same amount to $Q$ as
the powerlaw component.   
As we have verified in a series of test calculations, the EUV spectral
shape generally has no strong influence on the X-ray properties 
of the warm gas, its ionization state
is dominated by the X-ray part of the incident continuum (Komossa \& Fink 1997a).  
This also holds for the case of NGC 3227. 
(The re-calculation of the models to include an additional EUV bump was performed
to check the robustness of predictions made for the absorber-intrinsic emission and
absorption in lines in the optical and UV spectral region; cf. Sects 5.3.1, 5.3.2.)  
Furthermore, the {\em observed} UV continuum of NGC 3227 is unusually 
steep which seems to be due to strong  
reddening (Sect. 5.1.2), whereas the warm absorber most probably sees
the unreddened continuum. 
More explicitly, the mean Seyfert SED employed consists of an UV-EUV powerlaw 
of energy index $\alpha_{\rm uv-x}$=--1.4 (Kinney et al. 1991),
extending up to 0.1 keV, a mean continuum after Padovani \& Rafanelli (1988)
from the radio to the optical region with  
a break at
10$\mu$m and an energy index $\alpha$ = --2.5 $\lambda$-longwards, and an X-ray powerlaw 
that breaks into the
gamma-ray region at 100 keV. 
Since it turned out that the data do not allow to   
well constrain the underlying X-ray powerlaw index,  
it was usually fixed to (i) the canonical value of \G = --1.9 (used if not stated otherwise), 
that was found for NGC 3227 by George et al. (1990) in the \ginga observation, and (ii) 
\G = --1.6, the mean value derived by Ptak et al. (1994) using \asca data.     
   \begin{table*}                     
     \vspace{-0.5cm}
%     \begin{center}
     \caption{X-ray spectral fits to NGC 3227 (pl = powerlaw, disk = accretion disk model 
              as described in the text,  
                  wa = warm absorber).
       The errors are quoted at the 90\% confidence level. 
        }
     \label{fitres}
      \begin{tabular}{llccllccc}
      \hline
      \noalign{\smallskip}
        model & $N_{\rm H}$ & log $U$ & log $N_{\rm w}$ & log Norm$_{\rm pl}$
                            & $\Gamma_{\rm x}$ & $M_{\rm BH}$ & ${\dot M}/{\dot M_{\rm edd}}$ 
                            & $\chi^2_{\rm red} (d.o.f)$ \\
       \noalign{\smallskip}
%      \hline
      \noalign{\smallskip}
           & [10$^{21}$ cm$^{-2}$] & & [cm$^{-2}$] & & & [10$^4$ M$_{\odot}$] & & \\
       \noalign{\smallskip}
      \hline
      \hline
      \noalign{\smallskip}
 pl & 0.38$\pm$0.02 & - & - & --2.60$^{+0.05}_{-0.06}$$^{(1)}$
                & --1.19$\pm$0.11 & - & - & 1.67(18) \\
      \noalign{\smallskip}
      \hline
      \noalign{\smallskip}
 pl+disk & 0.70$\pm$0.16 & - & - & --2.60$^{+0.09}_{-0.12}$$^{(1)}$
                & --1.1$\pm$0.6 & 3.0$\pm$0.8 & 1.0$^{(3)}$ & 0.58(16) \\
      \noalign{\smallskip}
      \hline
      \noalign{\smallskip}
 wa & 0.55$\pm$0.13 & --1.03$\pm$0.10 & 21.49$\pm$0.13 & --4.24$\pm$0.02$^{(2)}$
                & --1.9$^{(3)}$ & - & - & 0.78(17) \\
      \noalign{\smallskip}
      \hline
      \noalign{\smallskip}
 dusty wa & 0.55$\pm$0.04 & --0.25$\pm$0.25 & 21.75$\pm$0.10 & --4.22$\pm$0.03$^{(2)}$
                & --1.9$^{(3)}$ & - & - & 0.69(17) \\
      \noalign{\smallskip}
      \hline
      \noalign{\smallskip}
  \end{tabular}

\noindent{\small $^{(1)}$ Normalization at 1 keV ~~~ $^{(2)}$ at 10 keV ~~~ $^{(3)}$ fixed
}
%%  \vspace{-0.4cm}
%   \end{center}
   \end{table*}

\subsubsection{Model results} 
Fitting the warm absorber model to the X-ray spectrum of 
NGC 3227, we find    
an ionization parameter of $\log U \simeq -1.0$ and a column density 
of the ionized material of $\log N_{\rm w} \simeq 21.5$, with 
$\chi^2$/dof = 13.3/17 (Table 1).
The mean observed X-ray flux for this model is $f$ = 2.15 $\times 10^{-11}$ erg/cm$^2$/s, 
corresponding to an intrinsic (0.1--2.4 keV) luminosity corrected for cold and warm absorption
of $L_{\rm x} = 0.8 \times 10^{42}$ erg/s.
The contribution to the \ros spectrum of 
emission and reflection from the warm material, 
calculated with the code {\em Cloudy} for a covering factor of the
ionized absorber of 0.5, is found to be negligible.    
The cold column,
$N_{\rm H} \simeq 0.55 \times 10^{21}$ cm$^{-2}$ 
is larger than the Galactic value
($N_{\rm H}^{gal} \simeq 0.22 \times 10^{21}$ cm$^{-2}$; derived upon
interpolation of the Dickey \& Lockman (1990) data),
but not as large as  
implied by emission-line reddening, as is further discussed in Sect. 5.1. 
Dust mixed with
the warm gas could supply the reddening without implying a corresponding cold column.

Since the presence of dust clearly modifies the resulting X-ray absorption structure
(Komossa \& Fink 1997a, 1997c), we re-calculated the models, now including dust with Galactic ISM
properties. (A density of the warm material of log $n_{\rm H}$ = 7 was used in
these models. Whereas usually the X-ray absorption does not strongly depend on density,
this value was chosen to ensure dust survival.)     
Applying the dusty absorber model to the data provides 
an excellent fit with $\chi^2$/dof = 11.7/17 (Fig. \ref{SEDx}). We find $\log U \simeq -0.25$ and
$\log N_{\rm w} \simeq 21.8$ and a cold column of 
$N_{\rm H} \simeq 0.55 \times 10^{21}$ cm$^{-2}$ for \G = --1.9.
(For comparison, the corresponding values for the \G = --1.6 model are:
$\log U \simeq -0.6$, $\log N_{\rm w} \simeq 21.5$, $N_{\rm H} \simeq 0.47 \times 10^{21}$ cm$^{-2}$,
and $\chi^2$/dof = 13.9/17.)   
 
The residuals from the best spectral fits, as compared to the single powerlaw
model, are displayed in Fig. \ref{SEDx}. The
unfolded X-ray spectrum is shown in Fig. \ref{wa_def}. 
No distinction between the dusty and dust-free absorber model is possible on the basis 
of the quality of the {\em X-ray} spectral fits (Table~1).  

\section {Temporal analysis}

\subsection {Flux variability}

%%\begin{figure}
%%\picplace{6cm}
  \begin{figure}[thbp]
      \vbox{\psfig{figure=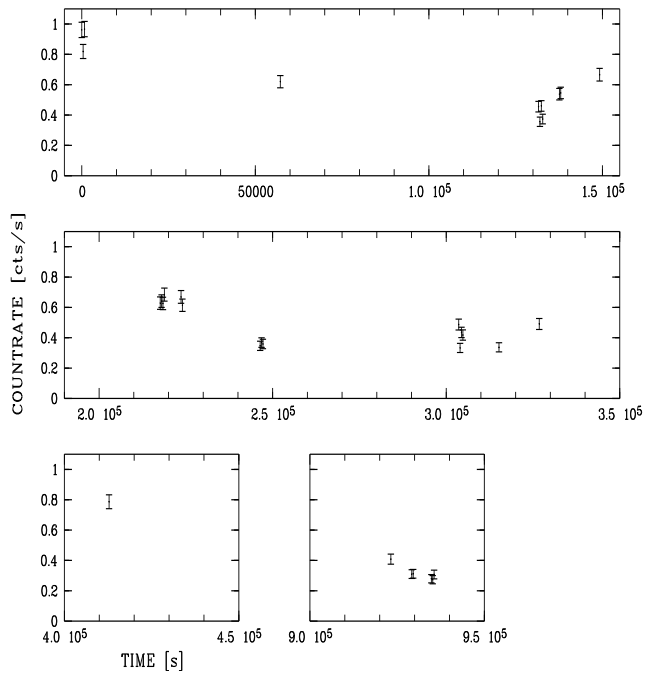,width=8.8cm,height=9.0cm,%
          bbllx=3.0cm,bblly=1.1cm,bburx=18.0cm,bbury=12.2cm,clip=}}\par
%      \vspace{-0.4cm}
 \caption[light]{X-ray lightcurve of NGC 3227 during the pointed observation,
binned to time intervals of 400 s. The time
is measured in seconds from the start of the observation.
   }
 \label{light}
\end{figure}

The X-ray lightcurve for the pointed observation is shown in Fig. \ref{light}.
With respect to the minimal resolved time interval of 400 s,
the observed count rate varies between 0.27 cts/s and 0.97 cts/s,
corresponding to an amplitude of variability of a factor of 3.5
(occuring within the total observing period of about 10.8 days).
The shortest resolved doubling timescale is 380 minutes 
and the strongest change within 
400 s is a factor of 1.5 drop in count rate.

Even stronger variability is detected between survey and pointed
observation (separated by $\sim$ 3 y),  
with a factor $\sim$ 8 variability in the mean count rates, and a factor of
$\sim$ 15 change, when comparing the survey
count rate with the maximum count rate of the pointed data.          
  \begin{figure}[thbp]
      \vbox{\psfig{figure=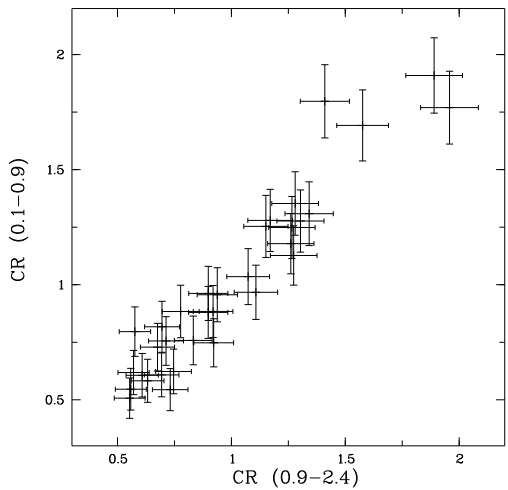,width=7.5cm,%
          bbllx=2.3cm,bblly=1.1cm,bburx=15.0cm,bbury=12.2cm,clip=}}\par
%      \vspace{-0.4cm}
 \caption[sh]{Count rate CR in the soft (0.1--0.9 keV) \ros energy band versus 
count rate in the hard (0.9--2.4 keV) band, each normalized to the mean count rate
in the corresponding band. There is correlated variability between both bands.
   }
 \label{sh}
\end{figure}
%----------------
  \begin{figure}[thbp]
      \vbox{\psfig{figure=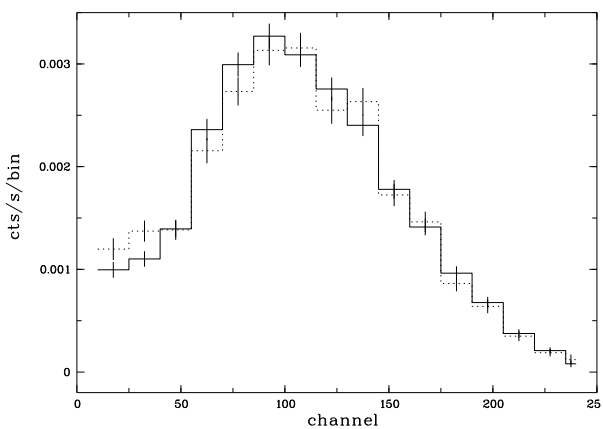,width=8.1cm,%
          bbllx=2.3cm,bblly=1.1cm,bburx=18.1cm,bbury=12.2cm,clip=}}\par
%      \vspace{-0.4cm}
 \caption[hl]{Count rate versus channel number in the range 0.11--2.4 keV: Low-state
data (dotted line) as compared to normalized high-state data (solid line).
 }
 \label{hl}
\end{figure}

\subsection{Spectral variability}

To check for correlated spectral changes in the hard and soft \ros energy range,
we have divided the data in a hard and soft band, 
with a dividing photon energy of 0.9 keV. 
This value was chosen to include most of the 
warm absorption features in the soft band, which extend throughout this whole
region for NGC 3227 (unlike in other objects, 
where the high-ionization oxygen and neon edges dominate). 
The soft band also includes the effects of the cold absorbing column, which consequently
cannot be disentangled from that of the ionized material (see below), 
whereas the hard band is dominated by the powerlaw continuum. 
We find correlated variability between both bands within the errors (Fig. \ref{sh}), 
implying that no strong spectral changes have taken place.  
 
To further constrain the spectral variability during the observation,
and to check for changes in the ionization
state of the warm absorber in more detail, the total observation was split (i) into individual 
subsets (referred to as `orbits') containing about 1000 to 2000 photons each, time-ordered, and merged in such a way,
that only photons of comparable source flux-state are combined, (ii) into only
two different subsets, `high-state' (more than 0.51 counts/s)
and `low-state'. The statistics are better in the latter approach, but it tends to smear out
trends if the spectral shape at a fixed flux state depends also on the history. 

(i) We find constant absorption structure, described by the ionization parameter $U$,
 within the errors 
throughout the observation, 
independent of variations in the      
intrinsic luminosity 
(log $U$ = --1.05, --1.04, --1.05, --1.01, --1.01, --1.00, --1.02 
for orbits 1 to 7, respectively and \G = --1.9). 
The cold and warm column densities have been fixed 
to the values determined for the total observation in this analysis.  
The same result holds for a flatter underlying
powerlaw and the dusty models. 
The constancy of the absorption structure allows to estimate an 
upper limit for the density of the warm absorber.  
The recombination timescale of the gas is given by 
$t_{\rm{rec}} \approx {n_{\rm{i}}\over n_{\rm{i+1}}}~n_{\rm{e}}^{-1}~\alpha_{\rm{i+1,i}}^{-1}$ 
(e.g. Krolik \& Kriss 1995), where $n_{\rm{i}}/n_{\rm{i+1}}$ is the ion abundance ratio of the major coolant,
$\alpha_{\rm{i+1,i}}$ the corresponding recombination rate coefficient (Shull \& Van Steenberg 1982), 
and $n_{\rm{e}}$ the electron density. 
For the best-fit warm absorber model (Table 1; values in brackets refer to the dusty model)
we find $n_{\rm{e}} \simeq 0.8 (0.2) \times 10^{11} t_{\rm rec}^{-1}$. No reaction 
of the warm material during the low-state at the end of the observation implies
$n_{\rm{e}} \approxlt 7 (2) \times 10^{6}$ cm$^{-3}$. No reaction during the total
observation, still using $t_{\rm{rec}}$ as an estimate, yields 
$n_{\rm{e}} \approxlt 9 (2.5) \times 10^{4}$ cm$^{-3}$.   
  
(ii) If $U$ is taken to be the only free parameter, the same result as for the study
of individual orbits is found.   
However, a more complicated situation with compensating changes of several
parameters cannot be excluded.  
When $N_{\rm w}$ is left as an additional free parameter, there 
is a slight trend for $U$ to be {\em lower} in the high-state data (which
is not the way one would expect the warm absorber to behave). 
When $U$, $N_{\rm w}$ and $N_{\rm H}$ are all left free, 
we find {\em higher} $U$ in the high-state data, but also  
{\em larger} $N_{\rm H}$ (reflecting the tendency shown in Fig. \ref{hl},
i.e. a deficiency of very low energy photons and an excess around the 
location of the oxygen absorption edges in the high-state).  
However, the fitted quantities come with larger errors, and within these, 
they are still consistent with being constant. 
These trends hold independent of the exact value of the underlying powerlaw index,
i.e. cannot be mimicked by variability of \G.  
They are only briefly mentioned here and will not be discussed further. 
In order to study short-timescale spectral variability in detail,
much better photon statistics are required.   
% -----------------
  \begin{figure}[thbp]
      \vbox{\psfig{figure=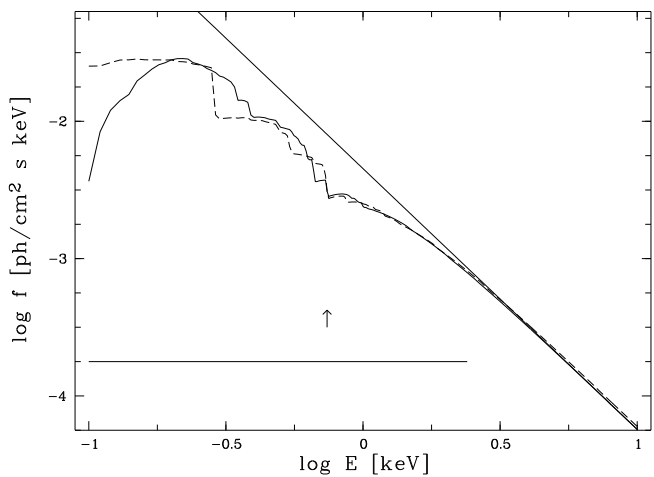,width=8.8cm,%
          bbllx=2.7cm,bblly=1.1cm,bburx=18.2cm,bbury=12.2cm,clip=}}\par
%      \vspace{-0.4cm}
 \caption[wa_def]{Warm-absorbed X-ray spectrum of NGC 3227 (dashed: dusty model)
between 0.1 and 10 keV, corrected for cold absorption. The straight line corresponds
to the unabsorbed powerlaw.  
The horizontal line at the bottom brackets the \ros energy range.  
 The arrow marks the position of the OVII edge. 
No iron edge is predicted
to arise from the warm material. 
}
 \label{wa_def}
\end{figure}

\section{Discussion}

\subsection{ Evidence for a {\em dusty} warm absorber}

\subsubsection{X-ray spectral fits} 

Two spectral models are found to be of similar success, one is a dust-free
warm absorber, the other one invokes dust mixed with the warm material. 

The presence of dust modifies the X-ray spectrum by
absorption and scattering (e.g. Martin 1970, Martin 1978, 
Voit 1991, Laor \& Draine 1993, Czerny et al. 1995).     
It also influences the conditions in the gas by 
several heating and cooling processes 
(e.g. Draine \& Salpeter 1979, Ferland \& Netzer 1979, Baldwin et al. 1991,
Netzer \& Laor 1993), 
like gas heating by grain photoionization and grain heating by 
collisions with gas particles.  
The clearest signature in the soft X-ray region is the presence of absorption edges that
are produced by inner-shell photoionization (e.g. Martin \& Rouleau 1991;
their Figs 4,5). The largest edge is that of carbon at an energy
of $\sim$280 eV (cf. our Fig. \ref{wa_def}); the shift of the edge energy due to 
solid state effects (Martin 1970) is only of the order of a few eV 
(e.g. Greaves et al. 1984).      

Due to the low abundance of dust as compared to the gas, and for typical cold 
Galactic columns, 
X-ray dust features are usually not detectable, since  
it is difficult to distinguish from gas-phase absorption
due to the same element if the gas is {\em cold}.  
However, since
{\em warm} gas is more transparent to very soft X-rays and typically of
larger column density, dust signatures get
more clearly visible. Still, the carbon edge itself cannot 
be resolved with \ros. Its presence clearly reduces the count rate
in the low-energy channels, though.   

Since, for relatively high values of $U$, dust effectively competes with the gas in the absorption of
photons (e.g. Laor \& Draine 1993),
the spectral model involving a dusty absorber requires a higher ionization parameter
than the dust-free one (Table 1). 
A slightly larger total column density of the warm absorber is found for the
dusty model, reflecting the depletion of gas-phase oxygen which is partly bound
in the dust.  

Based on the quality of X-ray spectral fits only, no distinction 
between the dust-free and dusty warm absorber is possible in the case of NGC 3227. 
Arguments in favor of the latter model are given below. These are derived from
a combination of the X-ray results with optical and UV properties of NGC 3227.   

\subsubsection{Comparison of X-ray, optical, and UV data}

As will be detailed in the following, there is evidence for a large amount of 
cold material  
along the line of sight to the active nucleus of NGC 3227 
which should imprint its presence on the soft X-ray
spectrum.  
The evidence comes from 
emission line reddening:   
The broad H$\alpha$/H$\beta$ ratio as measured by, e.g., Cohen (1983)
implies an extinction of 
$A_{\rm v} \simeq$ 1.4 (to derive $A_{\rm v}$, we have always assumed 
an unreddened value of H$\alpha$/H$\beta$=3.1; if there is a very strong
contribution from collisional excitation, the estimated amount of reddening 
is reduced). 
Due to heating by the radiation field of the central continuum source, dust
is not expected to survive in the BLR (e.g. Barvainis 1987, Netzer 1990, Laor \& Draine 1993).
Therefore, the material responsible for the broad line reddening should be located
along the line of sight, between observer and BLR. 
A dusty {\em outer} BLR dominating the line emission with no clouds along the line of sight
seems to be excluded by results from reverberation mapping.
These show the bulk of the emission to originate 
very near the center ($r \approx 17$ ld; Salamanca et al. 1994, Winge et al. 1995)
where dust will be destroyed. 

If the dust that extincts the broad lines 
is accompanied by an amount of gas as typically found in the
Galactic interstellar medium, a large cold absorbing column is expected to
show up in the soft X-ray spectrum.
For the warm absorber fit to the X-ray spectrum, we found a cold column of
$N_{\rm H} = 0.55 \times 10^{21}$ cm$^{-2}$, which is larger than the
Galactic value, $N_{\rm H}^{gal} \simeq 0.22 \times 10^{21}$ cm$^{-2}$, but
only corresponds to $A_{\rm v}$=0.3.   

Mixing dust with the {\em warm} gas could supply the reddening 
without implying a corresponding {\em cold} column.
For the dusty warm absorber description of NGC 3227, the column density
of the ionized gas is  
$N_{\rm w} \simeq 10^{21.75}$ cm$^{-2}$ (Table 1), 
resulting in $A_{\rm v} \simeq$ 3.1 which now somewhat overpredicts
the reddening. We suggest a slightly modified grain size distribution
with a higher abundance of larger dust grains,   
more similar to the large-R grains observed in the Orion nebula 
(e.g. Mathis \& Wallenhorst 1981, Baldwin et al.  
1991) that are characterized by less extinction in the optical and UV.
Mixing Orion-type dust within the warm material, instead of the standard 
MRN-type dust, the X-ray fit remains unchanged within the error bars.  
It also has to be kept in mind that the value of \G is not well constrained
by the present observations. For a flatter powerlaw (\G=--1.6; Ptak et al. 1994
instead of \G=--1.9; George et al. 1990) the column density of the
dusty warm material (Sect. 3.2.2) agrees with that estimated from broad line reddening
without requiring changes in the dust properties.   

Other evidence for reddening along the line of sight is also provided by
the rather steep observed UV continuum of NGC 3227.
Using the IUE measurements in Courvoisier \& Paltani (1992; LWP 2416 and SWP 21778),
we find a powerlaw between 2206\AA~ and 1253\AA~ of index
$\alpha_{\rm IUE} \simeq -2.9$, whereas values of $\alpha_{\rm uv} \simeq -1.4$
are typically found (e.g. Kinney et al. 1991). 
For comparison, extinction with $A_{\rm v}$ = 3.1
would change an intrinsic index of $\alpha_{\rm IUE} = -1.4$ to $\simeq -4.1$
(using the extinction curve as listed in Osterbrock (1989; his Table 7.2)). 
In the context of the dusty absorber scenario,
this again suggests 
 dust with reduced UV--extinction properties as mentioned
above.       

Several determinations of reddening towards NGC 3227 are listed in Table 2.
 
The cold absorbing column of $N_{\rm H} = 0.55 \times 10^{21}$ cm$^{-2}$ 
required for a successful spectral fit of a dusty absorber 
is still larger than the Galactic value. It is consistent with  Mundell et al. (1995b),
who find evidence for HI\,21cm absorption towards the continuum-nucleus of NGC 3227,
with an estimated column density of $N_{\rm H} \simeq 0.55 \times 10^{21}$ cm$^{-2}$.   

Finally, we note that even stronger reddening than that of the broad lines 
is occasionally reported for the narrow lines, which may, or may not, be linked to the presence
of the warm absorber.   
The narrow H$\alpha$/H$\beta$ ratio measured by, e.g., Cohen (1983)
implies an extinction of
$A_{\rm v} \simeq$ 1.2. 
Based on an observation taken at a different time
(Dec. 1993), Mundell et al. (1995a) report a much higher
value. They find
H$\alpha$/H$\beta$=15, which corresponds to $A_{\rm v} \simeq$ 4.5. 
Since under typical NLR conditions the emitted 
narrow Balmer lines should always be close to their case B recombination
(Brocklehurst 1971) value, a change in the H$\alpha$/H$\beta$ ratio  
has to be attributed to extinction by dust. 
However, dust {\em intrinsic} to the narrow line clouds
is not expected to in- and decrease by factors of several within years,
suggesting the extinction to be caused by an external absorber along the line
of sight. 
The expected accompanying very large cold absorbing column is not seen 
in the \ros X-ray spectrum. Both observations are not
simultaneous, but they are only separated by 6{$1\over2$} months. 
   
Dust internal to the {\em warm} gas could relax the strong time constraint
set by the quasi-simultaneous optical and X-ray observations; but would still require
a very special geometry (like the NLR emission being dominated by
clouds directly along the l.o.s. (consistent with Arribas \& Mediavilla 1994; their Table 2), 
located on the far side only (i.e. behind the nucleus as seen from the observer),
and absorbed by a fast-moving dusty cloud on the near side). 
The scenario is not discussed further here; a confirmation
of the puzzling optical observations would be valuable. 

\subsubsection{Alternatives and implications} 

An alternative possibility to explain the comparatively low cold column observed in 
the soft X-ray spectral region (as compared to the column 
inferred from broad line reddening) is
that we see mainly scattered soft X-rays. 
In that case, however, the X-ray emission 
of NGC 3227 should not be that rapidly
variable (Ptak et al. 1994; our Sect. 4.1).

An alternative, that cannot be excluded, is 
a factor of a few overabundant dust in the {\em cold} 
column, or a strong dominance of small grains, to explain the observed amount of BLR reddening. 
It requires stronger deviations of the dust properties from Galactic MRN dust,
as compared to the possibility of dust internal to the warm material, 
but certainly cannot be excluded with available data 
and we will include the dust-free warm absorber description of the data in
the following discussion. 

In this context, one should re-mention the early X-ray observations of NGC 3227,
for which the presence of large cold $N_{\rm H}$ was claimed. 
Since these had no spectral resolution 
in the soft energy range, and no warm absorber model was applied to the data,
the signature of the ionized material may well have mimicked a {\em large cold}
absorbing column (note that the warm absorber modifies 
the intrinsic X-ray spectrum well 
above 1 keV; Fig. \ref{wa_def}). 
To further test this, we have re-fitted the model points of the best-fit warm absorber
model, as well as the directly observed \ros X-ray spectrum, by a powerlaw model
with free {\em cold} absorbing column. We have cut off all data points below 1 keV. 
In this case, we find indeed a very large cold column of $N_{\rm H} \simeq 3 \times 10^{21}$ cm$^{-2}$,
similar to what was reported for early X-ray observations, but which is, in fact,  
mimicked by the recovering warm-absorbed
spectrum. 

To summarize, the extinction towards NGC 3227 seems to be caused
by either (i) a warm absorber with internal dust, 
or (ii) overabundant dust in the cold column.
We tentatively favor the dusty {\em warm} absorber since
it requires less strong deviations from typical Galactic dust properties.
 Simultaneous observations of emission lines, UV continuum, and soft X-ray
properties of NGC 3227 will be of high value.

   \begin{table} 
     \vspace{-0.5cm}
%     \begin{center}
     \caption{ Indicators of reddening/cold absorbing material towards NGC 3227.
               To derive $A_{\rm v}$ from emission lines, we have assumed an intrinsic intensity ratio
               of H$\alpha$/H$\beta$ = 3.1. 
               The conversion from $A_{\rm v}$ to $N_{\rm H}$ and vice versa
               was performed using a standard Galactic gas-to-dust ratio (e.g., Bohlin et al. 1978).
               } 
     \label{red}
      \begin{tabular}{llll}
      \hline
      \noalign{\smallskip}
  method & $A_{\rm v}$ & $N_{\rm H}$ & ref$^{(4)}$ \\
       \noalign{\smallskip}
         & [mag]       & [10$^{21}$/cm$^{2}$]   &            \\ 
       \noalign{\smallskip}
      \hline
      \hline
      \noalign{\smallskip}
  broad line reddening & 1.4  & 2.59 & [2]   \\
      \noalign{\smallskip}
                      & 1.4  & 2.59 & [6]   \\ 
      \noalign{\smallskip}
      \hline
      \noalign{\smallskip}
  narrow line reddening & 4.9  & 9.01 & [7] \\ 
      \noalign{\smallskip}
                        & 1.2  & 2.21 & [2]  \\
      \noalign{\smallskip}
                        & 1.7  & 3.13 & [3]  \\     
      \noalign{\smallskip}
                        & 4.5  & 8.30 & [1] \\    
      \noalign{\smallskip}
      \hline
      \noalign{\smallskip}
  10$\mu$ silicate feature & 7.5  & 13.8 & [4]  \\
      \noalign{\smallskip}
      \hline
      \hline 
      \noalign{\smallskip}
  HI\,21cm absorption & 0.3 & 0.55 $^{(1)}$ & [5]  \\  
      \noalign{\smallskip}
      \hline
      \noalign{\smallskip}
  X-ray absorption & 0.3 & 0.55 $^{(2)}$ & [8]  \\  
      \noalign{\smallskip}
             & 3.1  & 5.62 $^{(3)}$  & [8]  \\ 
      \noalign{\smallskip}
      \hline
      \noalign{\smallskip}  
  \end{tabular}

\noindent{\small  
$^{(1)}$ Intrinsic to NGC 3227.  
$^{(2)}$ Cold column density, including the Galactic contribution.
$^{(3)}$ Warm column density.   
$^{(4)}$~References: [1] Mundell et al. (1995a), [2] Cohen (1983), [3] Gonzales Delgado \& Perez (1997),
[4] Lebofsky \& Rieke (1979) and assumptions as in Reichert et al. (1985), 
[5] Mundell et al. (1995b), [6] Winge et al. (1995; 
their Table 4), [7] Rubin \& Ford (1968; the authors comment on a possible underestimate
of the H$\beta$ flux, and estimate the uncertainty 
to be less than about 50\%), [8] this work.  
 } 
   \end{table}
\subsubsection{{\em Dusty} warm absorbers in other active galaxies ?}  

Notwithstanding the cautious remarks of the previous section, we may address  
the question about the possibility of all warm absorbers being dusty. E.g., an identification
with the dusty transition region between BLR and NLR, proposed by Netzer \& Laor (1993),
or material evaporating from the molecular torus,    
would then come to mind.
 
The presence of dusty warm gas in the quasar  
IRAS 13349+2438 was suggested by Brandt et al. (1996)
through an analysis of the multi-wavelength properties of this object,   
and a similar model mentioned to possibly 
apply to MCG-6-30-15.
On the other hand, dust internal to the warm gas was found to be no viable description 
of the observed X-ray spectrum 
of the narrow-line Seyfert 1 galaxy NGC 4051 (Komossa \& Fink 1997a).
In general, 
we find the absorption structure to change towards a relative increase in the depths of absorption
edges of more lowly ionized species when including dust in the warm
absorber models (see Fig. 3 of Komossa \& Fink 1997c)
leading to an apparent {\em flattening} of the soft X-ray spectrum that can be partly traced back
to the presence of neutral metal ions bound in the dust that modify the X-ray absorption
structure (Sect. 5.1.1).  
One consequence is, that it is more difficult to reproduce apparently steep 
soft X-ray spectra (which require strong edges of highly ionized oxygen and neon to dominate)
as, e.g., observed in narrow-line Seyfert 1s. 

In order to assess the frequency of the presence of dust within warm absorbers,
objects with apparently {\em flatter} than usual powerlaw spectra are good candidates
for further study, in combination with evidence for excess reddening 
along the line of sight.
The clearest signature of the presence of dust is the Carbon edge at 0.28 keV.
To uniquely detect this feature, sensitivity at this energies and particularly
good spectral resolution is needed.  
The next mission that may achieve this
is AXAF.

\subsection{Further properties of the warm absorber}

{\em Location of the ionized material.}      \\ 
\indent (i) For the dust-free best fit ($\log U \simeq -1.0$), the density-scaled distance of the warm
absorber is $r \simeq 2 \times 10^{17}$/$\sqrt{n_{7}}$ cm,      
where $n = 10^{7} n_{7}$.  This compares to the typical
BLR radius of $r \simeq 17$ ld = $4 \times 10^{16}$ cm, as determined from reverberation
mapping (Salamanca et al. 1994, Winge et al. 1995). 

(ii) In case of dust internal to the warm absorber,  
the gas density has to be less than about $10^{7}$ cm$^{-3}$, i.e. 
$r \approxgt 8 \times 10^{16}$/$\sqrt{n_{7}}$ cm, 
to ensure dust survival.   
Constant ionization parameter $U$ throughout the observed low-state further implies
$n \approxlt 2 \times 10^6$ cm$^{-3}$ (Sect. 4.2); or even
$n \approxlt 2.5 \times 10^{4}$ cm$^{-3}$, based on
the assumption of constant $U$ during the total observation
(less certain, due to time gaps in the data).  
The ionized material should be located at least between BLR and NLR;
consistent with the radius of the BLR as determined from reverberation
mapping. 

% ___________________________________
  \begin{figure}[thbp]
      \vbox{\psfig{figure=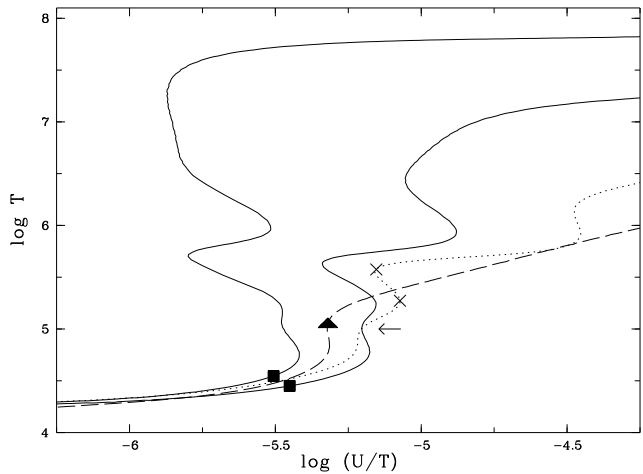,width=8.8cm,%
          bbllx=2.85cm,bblly=1.1cm,bburx=18.15cm,bbury=12.2cm,clip=}}\par
      \vspace{-0.4cm}
 \caption[stab]{Equilibrium temperature $T$ versus $U/T$.
The equilibrium curves are shown for different ionizing continua
and gas properties. 
Solid line: mean Seyfert continuum (as described in Sect. 3.2.1) 
with \G = --1.6 (left), \G = --1.9 (right);
dotted: observed multi-wavelength continuum of NGC 4051 with \G = --2.3, 
adopting solar abundances in all cases;
dashed: \G = --1.9, with depleted abundances and dust, as described in the text.
The fat symbols mark the location of the warm absorber in NGC 3227, for the
corresponding best-fit absorber models. For comparison, the position of the ionized
material in MCG-6-30-15 (arrow, solid line; Reynolds \& Fabian 1995)
and NGC 4051 at two different epochs (crosses; Komossa \& Fink 1997a,c)
is indicated.
}
 \label{stab}
\end{figure}

\paragraph{Thermal stability.} 
The thermal stability of the warm material
is addressed in Fig. \ref{stab}. All models were calculated with the code {\em Cloudy}. 
Between the low-temperature ($T \sim 10^4$ K) and high-temperature ($T \sim 10^{7-8}$ K) branch
of the equilibrium curve there
is an intermediate region of multi-valued behavior of $T$ in dependence
of $U/T$, i.e. pressure (e.g. Guilbert et al. 1983, their Fig. 1;
for a general discussion see also Krolik et al. 1981, Netzer 1990, Reynolds \& Fabian 1995).
It is this regime, where the warm absorber is expected to be located,
with its temperature typically found to lie around $T \sim 10^5$ K.
In case $T$ is multi-valued for constant $U/T$, and the gradient of
the equilibrium curve is positive, several phases may exist in
pressure balance.  
An analysis by Reynolds \& Fabian (1995) placed the warm absorber in
MCG-6-30-15 in a small stable regime within this intermediate region 
(its position
is marked by an arrow in Fig. \ref{stab}).
The ionized material in NGC 3227 is characterized by a comparatively
low ionization parameter. Its location in the $T$ versus $U/T$ diagram is
shown for the models that provide a successful description
of the X-ray spectrum. The shape of the equilibrium curve is clearly altered  
for the model including dust and correspondingly depleted metal abundances,
since both, dust and metals, affect the heating and cooling of the gas.
Consequently, due to this strong influence of several parameters on the shape of the
equilibrium curves (Fig. \ref{stab}), a detailed assessment of the 
stability and confinement of the ionized absorber in NGC 3227, and in general, 
is still
difficult in this approach.

\subsection{Predictions of warm-absorber intrinsic emission and absorption lines} 

\subsubsection{Line emission} 
Since warm absorbers are strongly matter-bounded and the electron temperature 
is high, their emissivity in emission
lines is usually expected to be
less than, e.g., that of the BLR. 
However, for the high values of $U$ typical for warm absorbers 
some optical--UV high-ionization lines can get very strong
(e.g. Netzer 1993, Hamann et al. 1995, \mbox{Komossa} \& Fink 1995, Reynolds \& Fabian 1995, 
Shields et al. 1995),
and one of the known high-ionization emission-line regions in active galaxies may
be identified with the ionized absorber. 
E.g., in many approaches to model the NLR, HeII$\lambda$4686 and other high-ionization
lines turned out to be systematically underpredicted.     
Shields et al. (1995) suggested a
matter-bounded BLR component to act as warm absorber.  

From a sample of individual objects chosen to investigate a possible warm-absorber
contribution to the observed emission lines, 
the one in NGC 3227 represents the low-ionization end of the known warm absorbers. 
Furthermore, the optical spectrum of NGC 3227   
shows broad wings in H$\alpha$ that do not follow continuum variations
(Salamanca et al. 1994). Interpreting large line-width as closeness to the nucleus
and high density, the non-variability indicates a higher-than-usual ionized component, 
that may be identified with the warm absorber.   

In the following we predict the emission lines expected to arise from 
the warm material, with its column density and ionization parameter
determined from the X-ray spectral fits. 
 
(i) Dust-free warm absorber:
Several emission lines are found to be rather strong,  
compared to the absorber-{\em intrinsic} 
H$\beta$ emission. The strongest are [FeX]$\lambda$6374/H$\beta$ = 1,  
 HeII$\lambda$4686/H$\beta$ = 1.1, in the optical region, and, e.g.,  
CIV$\lambda$1549/H$\beta$ = 170, NV$\lambda$1240/H$\beta$ = 15, OV$\lambda$1218/H$\beta$ = 18,
OVI$\lambda$1035/H$\beta$ = 74, NeVIII$\lambda$774/H$\beta$ = 0.3, in the UV -- EUV.  
However, the absorber-intrinsic luminosity in H$\beta$ 
itself is about a factor of 30 weaker than 
the mean de-reddened broad observed H$\beta$
emission, $L_{\rm H\beta} \simeq 10^{40.4}$ erg/s.
Scaling the line emission of the warm absorber correspondingly, 
to allow a judgement on the detectability of these features, results
in [FeX]$\lambda$6374$_{\rm{wa}}$/H$\beta_{\rm{obs}}$ = 0.03,
CIV$_{\rm{wa}}$/H$\beta_{\rm{obs}}$ = 6, etc.
CIV is the strongest predicted line. Its intensity also depends on density
(e.g., CIV$_{\rm{wa}}$/H$\beta_{\rm{obs}}$ = 4 for log $n_{\rm H} = 6$) and
on EUV continuum
shape (e.g., CIV$_{\rm{wa}}$/H$\beta_{\rm{obs}}$ = 3 for an additional
EUV black body as described in Sect. 3.2.1). 
To search for a warm-absorber contribution to individual lines, 
high-resolution, good quality spectra are needed.

(ii) Dusty absorber: In case of dust internal to the warm material, the overall
emissivity is reduced and there are no strong emission lines expected to be observed
in the IR to UV spectral region. The absorber-intrinsic ratio of, e.g., CIV/H$\beta$ is
weakened by a factor of more than 10, partially due to line destruction by grains.  

\subsubsection{UV absorption} 

A comparison of UV absorption lines and X-ray absorption properties of 
active galaxies was performed
by {\mbox Ulrich} (1988). 
Her analysis of an IUE spectrum of NGC 3227 in the range 
2000 -- 3200 \AA~revealed the presence of MgII$\lambda$2798 absorption
with an equivalent width of $\log W_{\lambda}$/$\lambda$ $\simeq$ --2.8.    
The X-ray warm absorber in NGC 3227 is of comparatively low ionization parameter.  
Nevertheless, the column density in MgII is very low, with $\log W_{\lambda}$/$\lambda$ = --5.7
(and considerably weaker for the dusty model or one with an additional
EUV black body). 
Predictions of UV absorption lines that arise from the ionized material 
are made for the more highly ionized species, with column densities 
of log $N_{\rm C^{3+}}$ = 17.5 (16.3) and log $N_{\rm N^{4+}}$ = 16.8 (15.7);
the numbers in brackets refer to the warm absorber model that includes dust.  
(For comparison, these numbers change as follows, if the EUV black body
component is again added to the incident continuum: 
log $N_{\rm C^{3+}}$ = 17.1 (16.1) and log$N_{\rm N^{4+}}$ = 16.7 (15.6).)  
The corresponding equivalent widths are then 
$\log W_{\lambda}$/$\lambda$ =~--2.9$^{+0.2}_{-0.4}$
(--~3.0$^{+0.2}_{-0.4}$)  
 for CIV$\lambda$1549, and $\log W_{\lambda}$/$\lambda$ = --3.0$^{+0.2}_{-0.4}$
(--3.1$^{+0.2}_{-0.4}$) for NV$\lambda$1240. 
The values are calculated for velocity parameters $b$ (Spitzer 1978) of 60 km/s,
100 km/s (for `+'), and 20 km/s (for `--').     
It will be worthwhile to search for these lines by obtaining high-quality UV spectra. 
For some examples of common UV -- X-ray warm absorbers see Mathur et al. (e.g., 1994, 1997),
Schartel et al. (1997). 

\section{Summary and  conclusions}
We have presented survey and pointed \ros PSPC observations of the Seyfert galaxy NGC 3227.

The temporal analysis of the data disclosed strong long-term variability,
with a maximum change of a factor $\sim$ 15 in count rate (within $\sim$ 3 years),
and smaller changes on the timescale of days.  

The spectral analysis, possible only for the pointing, revealed no strong
spectral variability.     
Two kinds of models provide a successful description of the soft X-ray spectrum,
a powerlaw with soft excess, and a warm absorber with or without internal dust.  
The first is not further discussed due to the unusually flat
inferred underlying powerlaw continuum.  

Both, the dust-free and the dusty warm absorber give excellent fits,
with values of the ionization parameter $U$ and warm column density $N_{\rm w}$ of
$\log U \simeq -1.0$, $\log N_{\rm w} \simeq 21.5$, and 
$\log U \simeq -0.25$, $\log N_{\rm w} \simeq 21.8$, respectively.
The clearest signature of the warm absorber with internal dust is the presence of 
a strong carbon edge at 0.28 keV that may be resolved with future X-ray missions.  
 
The absorber-intrinsic line emission and absorption in the optical--UV was studied.
UV absorption lines are predicted. Significant line emission is only found for
the dust-free description of the warm absorber, the strongest line being CIV$\lambda$1549.
However, no known emission-line region in NGC 3227 can be completely identified with the
warm absorber.

The {\em dusty} warm absorber model 
is shown to provide an   
explanation for the  
discrepancy
between the {\em low} cold absorbing column seen in the X-ray spectrum
and the {\em larger} one inferred from emission line reddening.
The density of the dusty warm material is constrained to at least
$n \approxlt 10^{7}$ cm$^{-3}$, to ensure dust survival,
which translates into a distance from the nucleus of
$r \approxgt 8 \times 10^{16}$/$\sqrt{n_{7}}$ cm and is 
a factor of several larger than the distance of the BLR as
determined from reverberation mapping. 

\begin{acknowledgements}
The \ros project is supported by the German Bundes\-mini\-ste\-rium
f\"ur Bildung, Wissenschaft, Forschung und Technologie (BMBF/DARA) and the Max-Planck-Society. 
We thank Gary Ferland for providing {\em{Cloudy}}, and Hartmut Schulz and the referee,
Hagai Netzer, for fruitful discussions and suggestions.      
This research has made use of the NASA/IPAC extragalactic database (NED)
which is operated by the Jet Propulsion Laboratory, Caltech,
under contract with the National Aeronautics and Space
Administration.
\end{acknowledgements}

\end{document}